\documentstyle[multicol,prl,aps,epsf]{revtex}

\begin{document}
\newcommand{\ftriangle}[1]{%
\setlength{\unitlength}{#1} 
\begin{picture}(5,3.2)
\put(4,0){\line(-1,0){4}}
\put(3.885,0.2){\line(-1,0){3.769}}\put(3.769,0.4){\line(-1,0){3.538}}
\put(3.654,0.6){\line(-1,0){3.307}}\put(3.538,0.8){\line(-1,0){3.076}}
\put(3.423,1.0){\line(-1,0){2.845}}\put(3.307,1.2){\line(-1,0){2.614}}
\put(3.192,1.4){\line(-1,0){2.383}}\put(3.076,1.6){\line(-1,0){2.152}}
\put(2.961,1.8){\line(-1,0){1.922}}\put(2.845,2.0){\line(-1,0){1.691}}
\put(2.730,2.2){\line(-1,0){1.460}}\put(2.614,2.4){\line(-1,0){1.229}}
\put(2.499,2.6){\line(-1,0){0.998}}\put(2.383,2.8){\line(-1,0){0.767}}
\put(2.326,2.9){\line(-1,0){0.651}}\put(2.268,3.0){\line(-1,0){0.534}}
\put(2.210,3.1){\line(-1,0){0.420}}
\end{picture}}
\draft
\title{Glass Transition Temperature and Dynamics of $\alpha$-Process in
Thin Polymer Films}
\author{K. Fukao\cite{A} and Y. Miyamoto}
\address{
Faculty of Integrated Human Studies,
Kyoto University, Kyoto 606-8501,
Japan 
}
\date{Received July 29, 1998} 
\maketitle
\begin{abstract}
The glass transition temperature $T_{\rm g}$ and the temperature 
$T_{\alpha}$ corresponding to the peak in the dielectric loss 
due to the $\alpha$-process
have been simultaneously determined as functions of film thickness~$d$ 
through dielectric measurements for thin films of polystyrene. 
A decrease of $T_{\rm g}$ was observed with decreasing film thickness,
while $T_{\alpha}$ was found to remain almost constant for $d>d_{\rm c}$ and 
decrease drastically for $d<d_{\rm c}$. Here, $d_{\rm c}$ is a critical 
thickness dependent on molecular weight. The thickness dependence of 
$T_{\rm g}$ is related to the distribution of relaxation times of the 
$\alpha$-process, not to the relaxation time itself. 
\end{abstract}

\vspace{-0.3cm}
\begin{multicols}{2}
Although many properties of glass transitions in glass-forming
materials have been clarified by recent experimental and 
theoretical works, the nature of glass transitions is not yet fully 
understood~\cite{Noncry}. Understanding the characteristic length scale of 
the dynamics of supercooled liquids near the glass transition 
is the most important issue in such studies.
Molecular dynamics simulations reveal the existence of significant 
large scale heterogeneity in particle displacements, so-called
{\it dynamical heterogeneity} in 
supercooled liquids~\cite{MD}. As the temperature decreases to 
$T_{\rm g}$, the dynamical heterogeneity grows. 
This phenomenon can be interpreted as the growth of 
the {\it cooperatively rearranging range} (CRR)
with decreasing temperature 
in the supercooled liquid state~\cite{Adam-Gibbs}. 
Multi-dimensional NMR~\cite{Spiess} and 
dielectric hole burning~\cite{Bohmer} reveal  
evidence of dynamical heterogeneity.
These topics concerning heterogeneity are closely  
related to the length scale of dynamics near glass transitions.

One of the most useful approaches to studying the length scale involved in  
dynamic glass transitions is to investigate the finite size 
effect on these transitions, i.e., to investigate the 
systems confined to nanopores~\cite{Kremer1} or thin 
films~\cite{Keddie1,DeMaggio,Forrest,Wallace}. 
In such systems, deviation from  
bulk properties is expected to appear if the system size is comparable
to the characteristic length scale. In particular, $T_{\rm g}$
of thin films has to this time been measured by 
several experimental techniques, including 
 ellipsometry~\cite{Keddie1}, positron annihilation~\cite{DeMaggio}, 
Brillouin light scattering~\cite{Forrest}, and X-ray 
reflectivity~\cite{Wallace}. Keddie {\it et al.} measured 
$T_{\rm g}$ of thin polymer films for the 
first time. For polystyrene films on hydrogen-passivated Si, $T_{\rm g}$
was found to decrease with decreasing film thickness $d$ for  
$d$$<$40nm~\cite{Keddie1}. In the case 
of freely standing polystyrene films, $T_{\rm g}$ decreases {\it much more 
rapidly} with decreasing film thickness~\cite{Forrest}. These results 
suggest that the
interaction between polymers and the substrate competes with surface
effects. This competition leads to the more gradual decrease of $T_{\rm g}$ 
in the former case. For a strong 
attractive interaction between polymers and the substrate, an increase in 
$T_{\rm g}$ was observed~\cite{Keddie1}. 

The dynamics related to the glass transition in thin films have
been investigated by many methods. Second harmonic generation reveals that 
the distribution of relaxation times broadens 
with decreasing film thickness, while the averaged relaxation time of 
the $\alpha$-process remains constant for the supported films of a 
random copolymer~\cite{Hall}. Atomic force microscopy~\cite{Kajiyama}
and photon correlation spectroscopy~\cite{Forrest2} studies show the 
existence of a mobile layer near the free surface of thin films of polystyrene.
However, it is not yet clear whether 
properties of the $\alpha$-process change together with the change
in $T_{\rm g}$ displayed as film thickness decreases.

In this Letter, we report on dielectric measurements made to determine the
$d$-dependence of $T_{\rm g}$ in atactic polystyrene (a-PS) by the 
temperature dependence of the electric capacitance, that is, to check 
whether $T_{\rm g}$ of thin films goes up or down compared with that 
of the bulk samples by another independent technique, and also we report
on an investigation of the relation between the $d$-dependence 
of $T_{\rm g}$ and that of the dynamics of the $\alpha$-process. 
We have successfully determined the $d$-dependence of $T_{\rm g}$.
These results agree well with those obtained by ellipsometry on
 supported PS films. 
The $d$-dependence of the width of dielectric loss peak of 
the $\alpha$-process is closely related to that of $T_{\rm g}$, while 
$T_{\alpha}$ corresponding to the peak in the dielectric 
loss due to the $\alpha$-process remains almost constant. We also 
found {\it the existence of a critical thickness} at which the dynamics of
the $\alpha$-process change drastically. We hope that these findings 
lead to a breakthrough in the investigation of the characteristic 
length scale of the glass transition.

Thin films of a-PS with various thicknesses 
from 4 nm to 489 nm were prepared using a spin-coat method from a
toluene solution of a-PS on Al-deposited slide glass. 
The thickness was controlled by changing the concentration of the solution. 
Two different a-PS's were used, one was purchased from Scientific 
Polymer Products, Inc. ($M_{\rm w}$=2.8$\times$10$^5$), and the other 
from Aldrich Co., Ltd. ($M_{\rm w}$=
1.8$\times$10$^6$, $M_{\rm w}/M_{\rm n}$=1.03). 
After annealing at 70$^{\circ}$C in the vacuum system for several days 
to remove solvents, Al was vacuum-deposited again to 
serve as an upper electrode. Heating cycles between room temperature 
and 110$^{\circ}$C ($>$$T_{\rm g}$) were carried out more than once before 
the dielectric measurements to relax the as-spun films and obtain 
reproducible results. Dielectric measurements were done using an LCR 
meter (HP4284A) for the frequency range from 20 Hz to 1MHz during 
heating (cooling) processes in which the temperature was changed at a 
rate of 2K/min.

\hspace*{-0.2cm}\begin{minipage}{8.5cm}
\begin{figure}
\epsfxsize=7.9cm 
%
\epsfbox{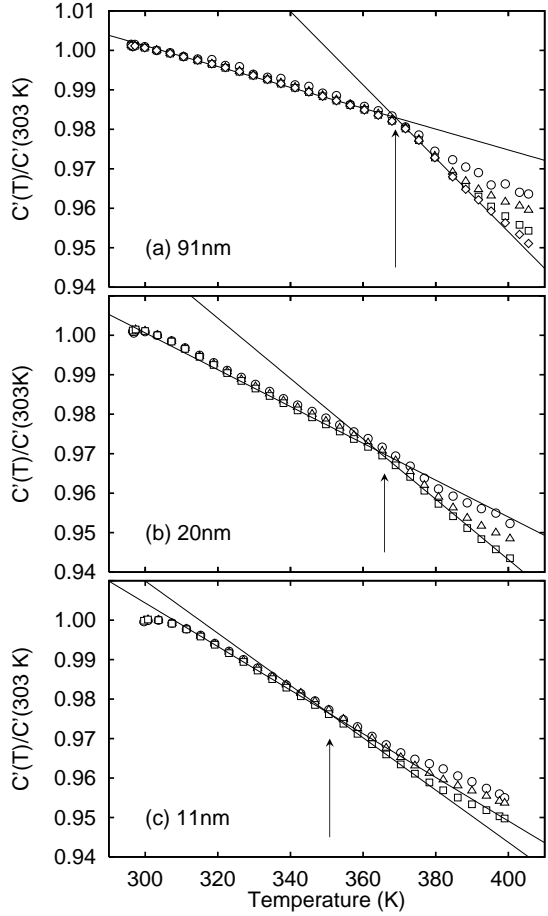}
\vspace{-4.8cm}
\caption{Temperature dependence of the capacitance normalized with
respect to the
values at 303K during the heating process for various frequencies 
($\circ$ corresponds to 100Hz, $\triangle$ to 1kHz, $\Box$ to 10kHz, 
and $\diamond$ to 100kHz) and three different thicknesses 
((a), $d$=$91$nm and 
$M_{\rm w}$=1.8$\times$10$^6$ ; (b), $d$=$20$nm and $M_{\rm w}$=
2.8$\times$10$^5$ ; (c), $d$=$11$nm and $M_{\rm w}$=1.8$\times$10$^6$). 
The solid lines were obtained by fitting the data at 10kHz to a linear 
function below and above $T_{\rm g}$. The arrows 
indicate the values of $T_{\rm g}$.
}
\label{fig:fig1}
\end{figure}
\vspace{0.1cm}
\end{minipage}

Film thickness was evaluated from the capacitance at room temperature
of as-prepared films by using the formula for the capacitance $C'$ 
of a flat-plate condenser, $C'$=$\epsilon'\epsilon_0S/d$, where 
$\epsilon'$ is the permittivity of a-PS, $\epsilon_0$ is the permittivity 
of the vacuum, $S$ is the area of the electrode ($S$=8.0mm$^2$), and $d$ 
is the thickness of the 
films. The value of $\epsilon'$  at room temperature is $\sim$2.8 for 
the bulk PS~\cite{Yano} and is assumed to be independent of $d$.
In the temperature range below that where the
 dielectric dispersions are observed,  the effect of dielectric 
dispersion can be neglected; i.e., here 
$\epsilon'(T)\approx\epsilon_{\infty}(T)$. 
In this case, the temperature dependence of $C'(T)$ comes only from that of
$\epsilon_{\infty}\epsilon_0S/d$, where $\epsilon_{\infty}$ is the 
permittivity in the high frequency limit. If we assume that the films are
constrained along the substrate surface, the thermal expansion along this 
surface can be neglected. Then the linear thermal expansion 
coefficient normal to the substrate $\alpha_{\rm n}$ is given 
by $\alpha_{\rm n}$=$(1+\nu)/(1-\nu)\alpha_{\infty}$, where $\alpha_{\infty}$
is the bulk linear coefficient of thermal expansion and $\nu$ is
Poisson's ratio~\cite{Wallace}. After taking account of the contributions from 
$\epsilon_{\infty}$ we obtain the temperature coefficient of capacitance
$\tilde\alpha$ as follows : $\tilde\alpha=-\frac{1}{C'(T_0)}\frac{dC'(T)}{dT}
\sim 2\alpha_{\rm n}$. We thus see that the temperature coefficient of $C'$ is 
proportional to $\alpha_{\rm n}$~\cite{Fukao1}.  Here, $T_0$ is a
standard temperature. It is therefore expected that the temperature 
coefficient should change at $T_{\rm g}$. 

\hspace*{-0.2cm}
\begin{minipage}{8.5cm}
\begin{figure}
\epsfxsize=9cm 
\hspace*{-0.6cm}\epsfbox{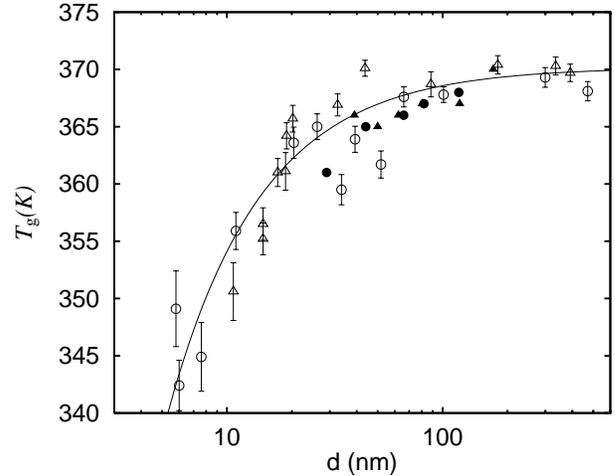}
\vspace{0.2cm}
\caption{Thickness dependence of $T_{\rm g}$ 
of a-PS films obtained during the heating process ($\circ$ corresponds
to $M_{\rm w}$=2.8$\times$10$^5$, and $\triangle$ to 
$M_{\rm w}$=1.8$\times$10$^6$). 
The values of $T_{\rm g}$ are determined as the crossover temperature 
between the straight lines characterizing $C'(T)$ at 10kHz below and 
above $T_{\rm g}$. The solid line was obtained from the equation 
$T_{\rm g}(d)=T_{\rm g}^{\infty}(1-a/d)$. 
The symbols  
$\bullet$ and \protect\ftriangle{0.5mm}
represent the data obtained for uncapped supported films of a-PS 
by Forrest {\it et al.} using ellipsometry for a-PS 
($M_{\rm w}$=7.67$\times$10$^5$ ($\bullet$), 2.2$\times$10$^6$ 
(\hspace*{-0.1cm}\protect\ftriangle{0.5mm}))~\protect\cite{Forrest}.
}
\label{fig:fig2}
\end{figure}

\vspace{0.1cm}
\end{minipage}

Figure 1 displays the temperature dependence of the capacitance, normalized 
with respect to the value at 303K during heating processes. 
In Fig.1(a) we can see that the values with thickness 91 nm 
for different frequencies can be reduced to a single straight line 
and decrease with increasing temperature for the temperature range 
from room temperature to approximately 370K.
At higher temperature the normalized capacitance decreases with increasing 
temperature more steeply than at lower
temperature. Here, the values for different frequencies can no longer be 
fitted by a single line, but are dispersed due to the appearance of the 
$\alpha$-process. For the temperature range shown in the
figure, however, there is almost no effect of the dispersion 
above 10kHz. Therefore, the temperature at which the slope of the
straight line of $C'(T)$ changes discontinuously can be determined 
unambiguously as the crossover temperature between the
straight line characterizing the lower temperature side and that 
characterizing the higher
temperature side for frequencies above 10kHz. This crossover temperature 
can be regarded as $T_{\rm g}$, because
the thermal expansion coefficient changes through the crossover 
temperature. The temperature coefficients of $C'$ 
obtained by fitting the 
data to the two lines are 2.6$\times$10$^{-4}$K$^{-1}$ 
for $T<T_{\rm g}$ and 8.7$\times$10$^{-4}$K$^{-1}$ for $T>T_{\rm
g}$ in the case $d$=91nm. These values agree well with those 
estimated from the values of 
$\alpha_{\infty}$ and $\nu$ in the literature~\cite{PolymerHand}.
As $d$ decreases, $T_{\rm g}$ also decreases, as shown in Figs.1(b) and (c). 

\hspace*{-0.2cm}
\begin{minipage}{8.5cm}
\begin{figure}
\epsfxsize=9.0cm 
\hspace*{-0.6cm}\epsfbox{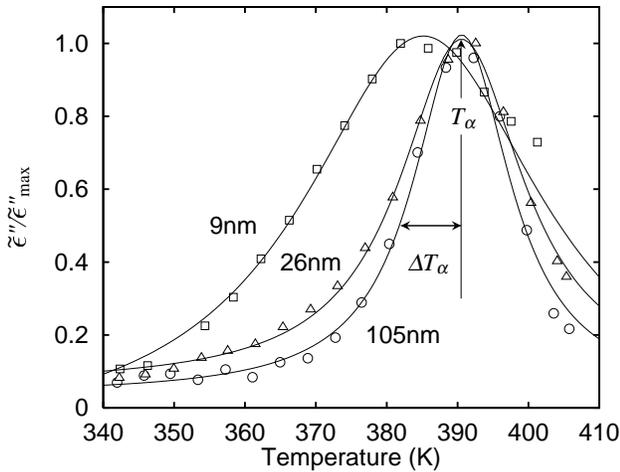}
\vspace{0.2cm}
\caption{Reduced dielectric loss as a function of temperature 
for various film thicknesses ($M_{\rm w}$=2.8$\times$10$^5$). 
The symbols $\protect\circ$ correspond to $d=105$ nm, 
$\triangle$ to $d=26$ nm, and $\Box$ to $d=9$ nm. The curves were
obtained by fitting the data to 
the equation $\epsilon''=\epsilon''_{\rm max}/(1+((T-T_{\alpha })/\Delta T_{\alpha })^2)$.}
\label{fig:fig3}
\end{figure}

\vspace{0.1cm}
\end{minipage}

Figure 2 displays the $d$-dependence of $T_{\rm g}$ for 
a-PS films determined as a crossover temperature at which the 
temperature coefficient of the capacitance at 10 kHz changes during 
heating process. 
When the films are thinner than 
about 100 nm, a decrease of $T_{\rm g}$ is observed. The value of
$T_{\rm g}$ for films of 6 nm thickness is lower by about 30K than 
that of films of 489 nm thickness. 
The dependence of $T_{\rm g}$ on $d$ can be expressed 
by a relation of the form 
$T_{\rm g}(d)=T_{\rm g}^{\infty}
\left(1-\frac{a}{d}\right)\approx T_{\rm g}^{\infty}\left(1+\frac{a}
{d}\right)^{-1},$
where $T_{\rm g}(d)$ is the measured glass transition temperature for a
 film of thickness $d$. The best fit parameters are $T_{\rm g}^{\infty}=370.2
\pm 0.4$~K and $a=0.43\pm0.03$~nm. Because these results agree well with 
those reported by Forrest {\it et al.}~\cite{Forrest}, it can be 
concluded that $T_{\rm g}$ of thin films has successfully be 
determined by measurements of electric capacitance for the first time. 
Here, we should note that $a$ is much smaller than the values of $d$ in 
our measurements. 
The value $a$ may be related to heterogeneous
distribution of mobility within films, as discussed later.

\hspace*{-0.2cm}
\begin{minipage}{8.5cm}
\begin{figure}
\epsfxsize=7.8cm 
\epsfbox{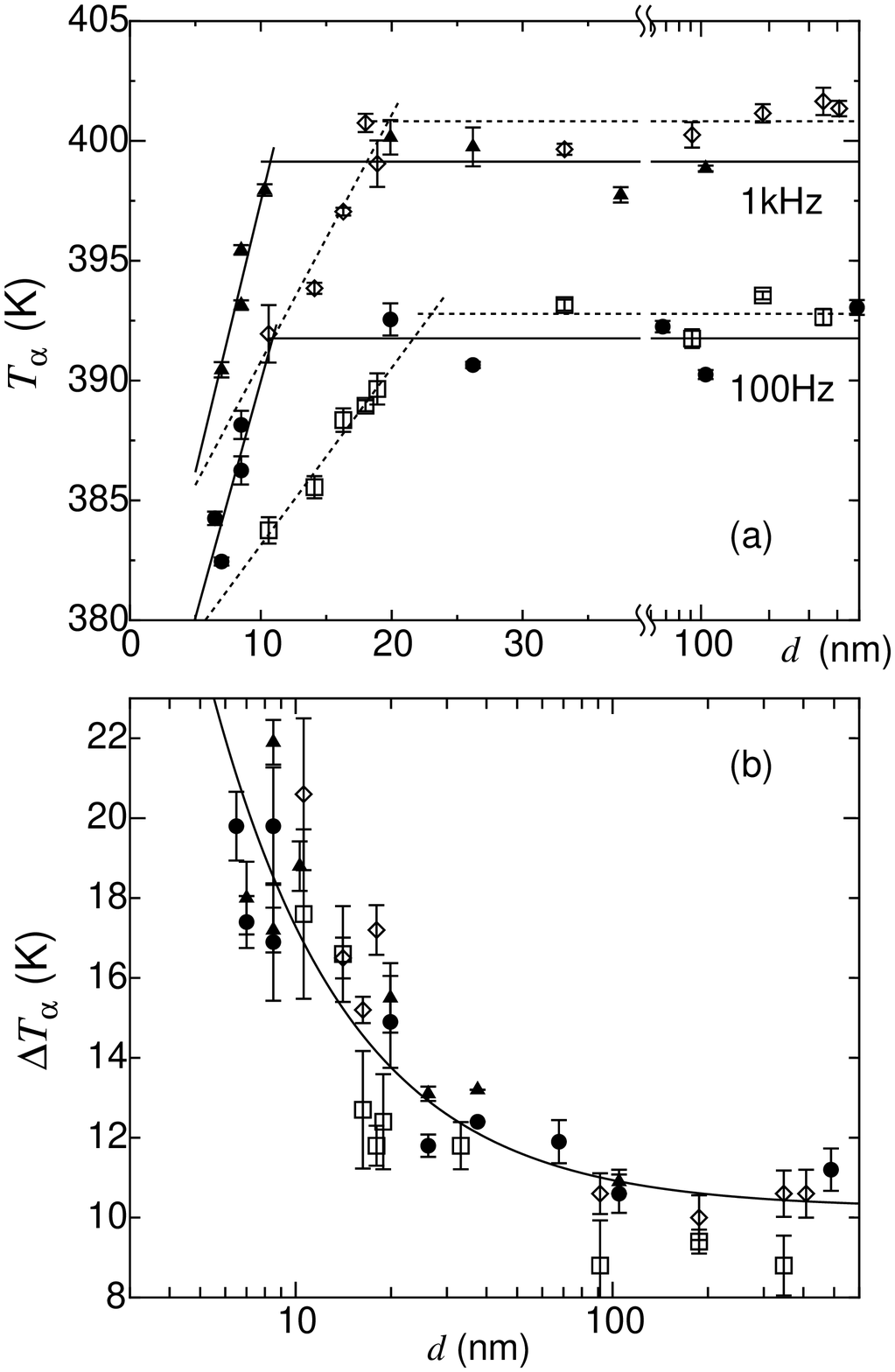}
\vspace{0.0cm}
\caption{(a) Thickness dependence of $T_{\alpha}$ and (b) $\Delta T_{\alpha}$
during the heating 
process at constant frequency. The symbol 
$\bullet$ corresponds to $f$=100Hz and $M_{\rm w}$=2.8$\times$10$^5$,%
\protect\ftriangle{0.5mm} to $f$=1kHz and $M_{\rm w}$=2.8$\times$10$^5$, 
$\Box$ to $f$=100Hz and $M_{\rm w}$=1.8$\times$10$^6$, and 
$\Diamond$ to $f$=1kHz and $M_{\rm w}$=1.8$\times$10$^6$. 
The solid lines in (a) were drawn for $M_{\rm w}$=2.8$\times$10$^5$ 
and the dotted lines, for $M_{\rm w}$=1.8$\times$10$^6$. 
These lines were drawn  by the equation 
$T_{\alpha}(d)= T_{\alpha}^{\infty}~\mbox{for}~d > d_{\rm c}$, 
$T_{\alpha}^{\infty}(1-(d-d_{\rm c})/\zeta)~\mbox{for}~d < d_{\rm c}$, 
where $T_{\alpha}^{\infty}$ and $\zeta$ are constants.
The curve in (b) was obtained from the equation $\Delta T_{\alpha}(d)=
\Delta T_{\alpha}^{\infty}(1+a'/d)$.
}
\label{fig:fig4}
\end{figure}

\vspace{0.1cm}
\end{minipage}

We now discuss how the dynamics of the $\alpha$-process change with  
decreasing $T_{\rm g}$ resulting from decreasing $d$. 
Figure 3 shows the reduced dielectric loss as a function of temperature 
at 100Hz in a-PS of thickness 9nm, 26 nm and 105 nm.
Above $T_{\rm g}$ the dielectric loss $\epsilon''$ at constant frequency 
displays an anomalous increase with temperature due to the
$\alpha$-process, and it possesses a maximum at the temperature 
$T_{\alpha}$. 
The value of $T_{\alpha}$ and the width of the $\alpha$-peak,  
$\Delta T_{\alpha}$, also depend on $d$, as shown in Fig.3. Here, 
$\Delta T_{\alpha}$ is defined as the temperature difference 
between the temperature $T_{\alpha}$ and the lower temperature 
at which $\epsilon ''$ is half its peak value. 
As shown in Fig.4(b), the width $\Delta T_{\alpha}$
 begins to increase at about 100 nm, and continues to increase monotonically 
with decreasing $d$. The $d$-dependence of 
$\Delta T_{\alpha}$ can be expressed by the equation 
$\Delta T_{\alpha}(d)=\Delta T_{\alpha}^{\infty}
(1+\frac{a'}{d})$, where $a'$=6.9$\pm$0.6 nm and 
$\Delta T_{\alpha}^{\infty}$=10.2$\pm$0.4 K. Comparing the $d$-dependence of 
$\Delta T_{\alpha}$ with that of $T_{\rm g}$ (Fig.2), we find that
the lowering of $T_{\rm g}$ is directly correlated with the 
broadening of the $\alpha$-peak as follow: 
$\delta(T_{\rm g}(d))/T_{\rm g}^{\infty}$=
$-6.0\times 10^{-2} \delta(\Delta T_{\alpha}(d))/
\Delta T_{\alpha}^{\infty}$, 
where $\delta T_{\rm g}(d)$=$T_{\rm g}(d)-T_{\rm g}^{\infty}$ and 
$\delta(\Delta T_{\alpha}(d))$=$\Delta T_{\alpha}(d)-\Delta 
T_{\alpha}^{\infty}$.
In other words, the broadening of the distribution of relaxation times 
for the $\alpha$-process is closely correlated to the reduction of $T_{\rm g}$. 

Contrastingly, Fig.4(a) shows that $T_{\alpha}$ remains 
almost {\it constant} as $d$ is decreased, down to a critical
thickness $d_{\rm c}$, at which point it begins to {\it decrease linearly}
with $d$. The values of $d_{\rm c}$ clearly depend on the molecular
weight ($M_{\rm w}$) of a-PS: $d_{\rm c}$=11 nm 
for $M_{\rm w}$=2.8$\times$10$^5$ and
$d_{\rm c}$=20$\sim$23 nm for $M_{\rm w}$=1.8$\times$10$^6$. 
These values seem to be related to the radius of gyration
of the bulk polymer coil ($R_{\rm g}$=0.028$\times$$\sqrt{M}$
(nm)~\cite{DeMaggio}):
$R_{\rm g}$=15 nm for $M_{\rm w}$=2.8$\times$10$^5$ and 38 nm for
 $M_{\rm w}$=1.8$\times$10$^6$. 
The $M_{\rm w}$- and $d$-dependences of $T_{\alpha}$ are quite different   
from those of $T_{\rm g}$ and $\Delta T_{\alpha}$ in the present and previous  
measurements on supported PS films~\cite{Keddie1,Forrest}, or 
rather, they are similar to that of $T_{\rm g}$ for freely standing films of 
a-PS~\cite{Forrest}.

In case of thin polymer films supported on substrate, not only the
 surface effects but also the interaction between the substrate and
 films strongly affects the dynamics and the glass transition of the 
thin films. A {\it three-layer model} was introduced in order to 
explain such surface and interfacial effects~\cite{DeMaggio}. 
In this model it is
 assummed that within a thin film there are a mobile layer and an
 immobile layer in addition to a bulk-like layer. The $d$-dependence 
of $\alpha_{\rm n}$ observed in our measurements also supports the 
model~\cite{Fukao1}. 
On the basis of this model, the existence of mobile and immobile 
layers with constant layer thicknesses broadens the distribution of
relaxation times of the $\alpha$-process with decreasing $d$, $i.e.$, 
$\Delta T_{\alpha}$ is increased. Because $T_{\rm g}$ can be regarded 
as the temperature at which the anomalous increase in $\epsilon''$ 
begins, $T_{\rm g}$ decreases with increasing $\Delta T_{\alpha}$.
On the other hand, the temperature $T_{\alpha}$ remains constant,
because $T_{\alpha}$ is the temperature at which $\epsilon''$ has 
a maximum. 
If $d$ reaches a critical thickness $d_{\rm c}$, the thickness of 
the bulk-like layer becomes comparable to the characteristic length 
scale of the $\alpha$-process and, as a result, the dynamics 
change drastically. For $d$$<$$d_{\rm c}$, $T_{\alpha}$ decreases or
 increases depending on whether contributions from the mobile layer 
are stronger than those from the immobile layer.
In the present case, $T_{\alpha}$ happens to decrease drastically.

In this Letter, three different length scales $a$, $a'$ and $d_{\rm c}$ 
were extracted from the dielectric measurements. 
From the above discussions, $d_{\rm c}$ is expected to be related to 
the characteristic length scale $\xi$ for the $\alpha$-process of bulk
samples.  
In this case it is difficult to evaluate the exact value of $\xi$   
because surface and interfacial effects and the molecular weight
dependence of $d_{\rm c}$ should be taken into account~\cite{DeMaggio,Fukao1}. 
However,  $d_c$ can at least be regarded as the upper limit of $\xi$. 

Recently, Forrest {\it et al.}\cite{Forrest2} have obtained the 
$\alpha$-relaxation data with a characteristic time scale 
$\langle\tau\rangle$$\sim$2$\times$ 10$^{-4}$s by a quartz crystal 
microbalance technique for supported PS films with SiC particles. 
It was reported that `$T_{\rm max}$', which can be compared with $T_{\alpha}$, 
has a $d$-dependence similar to that of $T_{\rm g}$. 
The $d$-dependence of $T_{\rm max}$ observed in their measurements 
seems to be different from that in our measurements.
This may come from the difference between dielectric relaxation dynamics 
and mechanical relaxation dynamics.

In summary, we have confirmed 
the reduction of $T_{\rm g}$ with decreasing $d$
for capped thin films of a-PS by dielectric measurements. 
The $d$-dependence of $T_{\rm g}$ is directly correlated to that 
of the width of the peak due to the $\alpha$-process in the temperature 
domain, $i.e.$, the distribution of relaxation times of the 
$\alpha$-process. The dynamics of the $\alpha$-process change 
drastically for the films with thickness less than a critical 
thickness.

The work was partly supported by a Grant-in-Aid from the Ministry 
of Education, Science, Sports and Culture of Japan.



\end{multicols}

\begin{references}
\vspace{-1.7cm}
\bibitem[\dagger]{A}
Electronic address: fukao@phys.h.kyoto-u.ac.jp 
\bibitem{Noncry} ``Proceedings of the 3rd International Discussion 
Meeting on Relaxations in Complex Systems'', J. Non-Cryst. Solids. 
(1998) (to be published).
\bibitem{MD} T. Muranaka and Y. Hiwatari, Phys. Rev. {\bf E51}, R2735, (1995); 
R. Yamamoto and A. Onuki, J. Phys. Soc. Jpn., {\bf 66}, 2545 (1997); 
W. Kob {\it et al.}, 
 Phys. Rev. Lett. {\bf 79}, 2827 (1997).
\bibitem{Adam-Gibbs}
G. Adam and J.H. Gibbs, J. Chem. Phys. {\bf 43}, 139 (1965).
\bibitem{Spiess}
K. Schmidt-Rohr and H.W. Spiess, Phys. Rev. Lett. {\bf 66}, 3020 (1991).
\bibitem{Bohmer}
B. Schiener {\it et al.}, 
{\bf 274}, 752, (1996).
\bibitem{Kremer1}
J. Sch\"uller {\it et al.}, 
Phys. Rev. Lett. {\bf 73}, 2224 (1994);
M.~Arndt {\it et al.}, 
Phys. Rev. Lett. {\bf 79}, 2077 (1997); 
G. Barut {\it et al.}, 
Phys. Rev. Lett. {\bf 80}, 3543 (1998).
\bibitem{Keddie1} J.L. Keddie, R.A.L. Jones, and R.A. Cory, Europhys. Lett.,
{\bf 27}, 59 (1994); J.L. Keddie and  R.A.L. Jones, Faraday Discuss. 
{\bf 98}, 219 (1994).
\bibitem{DeMaggio}
G.B. DeMaggio {\it et al.}, 
Phys. Rev. Lett. {\bf 78}, 1524 (1997).
\bibitem{Forrest} J.A. Forrest {\it et al.}, 
Phys. Rev. Lett. {\bf 77}, 2002 (1996); J.A. Forrest,
 K. Dalnoki-Veress, and J.R. Dutcher, Phys. Rev. {\bf E56}, 5705 (1997).
\bibitem{Wallace}
W.E. Wallace, J.H. van Zanten, and W.L. Wu, Phys. Rev. {\bf E52}, R3329 (1995).
\bibitem{Hall}
D.B. Hall, J.C. Hooker, and J.M. Torkelson, Macromolecules, {\bf 30},
 667 (1997).
\bibitem{Kajiyama}
T. Kajiyama, K. Tanaka, and A. Takahara, Polymer, {\bf 39}, 4665 (1998).
\bibitem{Forrest2}
J.A. Forrest {\it et al.}, 
Phys. Rev. {\bf E58}, R1226 (1998).
\bibitem{Yano}
O.~Yano and Y.~Wada, J. Polym. Sci.: Part A-2, {\bf 9}, 669 (1971).
\bibitem{Fukao1}
K. Fukao and Y. Miyamoto, (in preparation).
\bibitem{PolymerHand}
{\it Polymer Handbook}, 3rd ed., edited by J. Brandrup and E.H. Immergut 
 (John Wiley, New York, 1989).
\end{references}
\end{document}